\documentclass[fleqn,usenatbib]{mnras}

\usepackage{newtxtext,newtxmath}
\usepackage[T1]{fontenc}

\DeclareRobustCommand{\VAN}[3]{#2}
\let\VANthebibliography\thebibliography
\def\thebibliography{\DeclareRobustCommand{\VAN}[3]{##3}\VANthebibliography}

\usepackage{graphicx}	
\usepackage{amsmath}	
\usepackage{gensymb}	
\usepackage{textcomp}
\usepackage{hyperref}
\usepackage{mathtools}
\usepackage{subfig}
\usepackage[flushleft]{threeparttable}

\graphicspath{{./figures/}}

\title[Two New DAHe White Dwarfs]{Two New White Dwarfs With Variable Magnetic Balmer Emission Lines}

\author[Joshua S. Reding et al.]{
Joshua S. Reding$^{1}$\thanks{E-mail: jsreding@unc.edu}, 
J. J. Hermes$^{2}$, 
J. C. Clemens$^{1}$, 
R. J. Hegedus$^{1}$, 
B. C. Kaiser$^{1}$ 
\\
$^{1}$Department of Physics and Astronomy, University of North Carolina at Chapel Hill, Chapel Hill, NC 27599, USA\\
$^{2}$Department of Astronomy, Boston University, 725 Commonwealth Ave., Boston, MA 02215, USA}

\date{Accepted XXX. Received YYY; in original form ZZZ}

\pubyear{2023}

\begin{document}
\label{firstpage}
\pagerange{\pageref{firstpage}--\pageref{lastpage}}
\maketitle

\begin{abstract}
We report the discovery of two apparently isolated stellar remnants that exhibit rotationally modulated magnetic Balmer emission, adding to the emerging DAHe class of white dwarf stars. While the previously discovered members of this class show Zeeman-split triplet emission features corresponding to single magnetic field strengths, these two new objects exhibit significant fluctuations in their apparent magnetic field strengths with variability phase. The Zeeman-split hydrogen emission lines in LP $705{-}64$ broaden from $9.4$ MG to $22.2$ MG over an apparent spin period of $72.629$ minutes. Similarly, WD J$143019.29{-}562358.33$ varies from $5.8$ MG to $8.9$ MG over its apparent $86.394$-minute rotation period. This brings the DAHe class of white dwarfs to at least five objects, all with effective temperatures within $500$ K of $8000$ K and masses ranging from $0.65{-}0.83\,M_{\odot}$.
\end{abstract}


\section{Introduction} \label{sec:intro}
\par White dwarf stars are typically photometrically stable objects; $97\%$ of those observed by the late \textit{Kepler} Space Telescope between its original mission and \textit{K2} Campaign 8 are apparently non-variable to within $1\%$ in the \textit{Kepler} filter bandpass \citep{Howell14, Hermes17b}. The remaining $3\%$ host a wide variety of variability mechanisms including pulsations \citep{Warner72, Winget81}, magnetic spots \citep{Maoz15}, and interactions with binary companions, planets, and planetary debris \citep{Vanderburg15, Hallakoun18, Vanderbosch20}. This variable sample provides a means to understand stellar activity and evolution, both intrinsic (e.g. internal structure and dynamics) and extrinsic (e.g. planetary system evolution and interactions with host stars).
\par Some particularly enigmatic variable white dwarfs stand out from this sample as evading explanation. Among these are the growing class of DAHe white dwarfs: apparently isolated stars whose spectra are characterized by magnetically split (DH) hydrogen Balmer (DA) emission (De), which also exhibit both photometric variability and corresponding time-dependent variations in their Balmer features. The first object discovered in this class was GD 356, and it remained the only member for 35 years \citep{Greenstein85}. Across these three-and-a-half decades, astronomers studied GD 356 and speculated as to the source of the emission despite there being no apparent companion to feed it.
\par The prevailing model for most of this time involved a conducting planet orbiting through the stellar magnetosphere, inducing an electromotive force which excites the stellar atmosphere into emission, in a unipolar inductor configuration akin to that which is active in the Jupiter-Io system \citep{Li98, Goldreich69}. It was further proposed that this planet could have formed from material cast off in a double white dwarf merger, similar to how planets are hypothesized to form around millisecond pulsars \citep{Wickramasinghe10, Podsiadlowski91}. However, GD 356 is only a low-amplitude variable ($P_{\rm rot}\!\approx\!115$ min) due to its rotation axis orientation never moving its emission region fully out of our line-of-sight, so behavior exhibited elsewhere on the stellar surface cannot be observed to provide additional information \citep{Brinkworth04, Walters21}.
\par A second discovery finally established the DAHe class with the identification of SDSS J$1252{-}0234$ \citep{Reding20}, which presents significant (${\sim}5\%$) photometric variability in SDSS-\textit{g} on a dominant period of $5.3$ minutes. The Balmer features in SDSS J$1252{-}0234$ (particularly H$\beta$) also transition on this photometric period from moderately broadened absorption at photometric maximum to Zeeman-split triplet emission at photometric minimum, which confirms that the emission region is localized on the stellar surface to magnetic spots. The rapid rotation is anomalous compared to typical white dwarf rotation periods of $0.5{-}2.2$\,days \citep{Hermes17c}, which might indicate that the object formed from a previous stellar merger \citep{Ferrario97, Tout08, Nordhaus11}. \citet{Gaensicke20} then discovered a third object, SDSS J1219+4715, which bears more of a resemblance to GD 356 with its Balmer emission never fully disappearing, but rotates on a slower timescale ($P\!\approx\!15.3$ hr).
\par In addition to their mysterious behavior, the DAHe white dwarfs also exhibit a remarkable uniformity in their physical characteristics. All three have masses near the white dwarf population average ($0.62\,M_\odot$; \citealp{GenestBeaulieu19}), effective temperatures $7500\text{ K}\!<\!T_\text{eff}\!<\!8500\text{ K}$, and mega-gauss magnetic fields ($B_\text{GD356}\!=\!11$ MG, \citealp{Greenstein85}; $B_\text{J1252}\!=\!5$ MG,  \citealp{Reding20}; $B_\text{J1219}=\!18$ MG, \citealp{Gaensicke20}). This homogeneity, and the non-detection of a planetary companion to GD 356 with targeted study, suggest that the emission behavior may in fact be intrinsic to white dwarfs at this evolutionary phase \citep{Walters21}. The recent discovery of a similar apparently isolated white dwarf with variable emission, but yet undetectable magnetism, further confounds the nature of this mechanism (SDSS J$041246.85{+}754942.26$; \citealp{Tremblay20}).
\par Here we announce the discovery of two new DAHe white dwarfs, LP $705{-}64$ (\textit{Gaia} $G\!=\!16.9$\,mag) and WD J$143019.29{-}562358.33$ ($G\!=\!17.4$\,mag; henceforth J$1430$), which each present a unique twist on the established Zeeman-split triplet emission seen in the previous three DAHe. LP $705{-}64$ shows two different emission poles in its spectral variability, with one prominently featuring the classical Zeeman triplet emission at H$\alpha$ and H$\beta$ like in GD 356, before transitioning to reveal significantly broader Zeeman emission measurable only at H$\alpha$. J$1430$ shows a pole of Zeeman-split triplet absorption in H$\beta$, which is filled asymmetrically by broader triplet emission half a rotation cycle later, while H$\alpha$ simultaneously reveals fainter triplet emission across the same transition. Both maintain the other established similarities to the known members of the DAHe class, including in temperature, mass, magnetic field strength, and location in observational Hertzsprung-Russell diagrams.
\par We describe our survey strategy which uncovered these objects and the corresponding observations in Section~\ref{sec:obs}, and follow with a description of our analysis in Section~\ref{sec:nlss}. We then discuss the context and broader implications of these objects, and summarize our conclusions in Section~\ref{sec:conc}.

{\renewcommand{\arraystretch}{1}
\begin{table}
  \centering
  \caption{\textit{Gaia} DR3 astrometric parameters and photometry, and estimated $T_\text{eff}$, $\log g$, and mass for LP $705{-}64$ and J$1430$ using H-atmosphere white dwarf models \citep{GentileFusillo21b, Kowalski06, Tremblay11}.\label{table:phot}}
  \begin{tabular*}{0.82\columnwidth}{l | r r}
   \hline
   Parameter & LP $705{-}64$ & J$1430$\\
   \hline
   RA (deg, J2016.0) & $8.8044$ & $217.5803$ \\
   Dec (deg, J2016.0) & $-12.4198$ & $-56.3995$ \\
   $\varpi$ (mas) & $18.986\pm0.088$ & $15.075\pm0.097$ \\
   $d$ (pc) & $52.67\pm0.24$ & $66.33\pm0.43$ \\
   $\mu_\alpha$ (mas yr$^{-1}$) & $148.58\pm0.09$ & $-14.79\pm0.07$ \\
   $\mu_\delta$ (mas yr$^{-1}$) & $-150.61\pm0.08$ & $4.09\pm0.07$ \\
   $v_\text{tan}$ (km s$^{-1}$) & $52.89\pm0.28$ & $4.83\pm0.06$ \\
   \hline
   $G$ & $16.888\pm0.003$ & $17.407\pm0.003$\\
   $G_\text{BP}$ & $16.989\pm0.011$ & $17.529\pm0.008$\\
   $G_\text{RP}$ & $16.675\pm0.010$ & $17.220\pm0.009$\\
   \hline
   $P_\text{WD}$ & $0.993$ & $0.993$ \\
   $T_\text{eff}$ (K) & $8440\pm200$ & $8500\pm170$\\
   $\log g$ & $8.34\pm0.05$ & $8.37\pm0.05$\\
   Mass ($M_\odot$) & $0.81\pm0.04$ & $0.83\pm0.03$\\
   \hline
  \end{tabular*}
\end{table}
}

\section{Survey Strategy and Observations} \label{sec:obs}

\subsection{\textsc{VARINDEX} Survey and \textit{Gaia} Archival Data}
\par
\begin{figure}
  \centerline{\includegraphics[width=1.0\columnwidth]{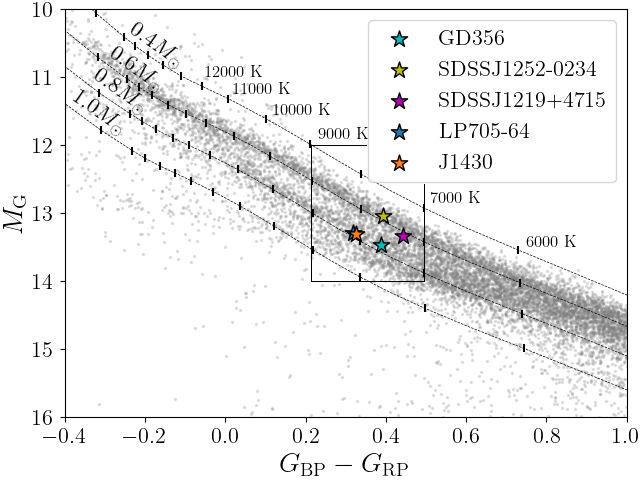}}
  \caption{\textit{Gaia} Hertzsprung-Russell diagram of white dwarfs within $100$ pc of the Sun in grey along with all DAHe objects discovered to date, \textsc{VARINDEX} survey window, and thick H layer mass tracks from \citet{Bedard20}. LP $705{-}64$ and J$1430$ have nearly identical locations and maintain the narrow DAHe parameter space.}
  \label{fig:GaiaHRD}
 \end{figure}

 We discovered the unusual activity in LP $705{-}64$ (\textit{Gaia} DR3 $2375576682347401216$) and J$1430$ (\textit{Gaia} DR3 $5892465542676716544$) using a survey strategy specifically formulated to identify likely DAHe candidates from the broader white dwarf population. We used the \textit{Gaia} DR2 \textsc{VARINDEX} metric, whose calculation is described in \citet{Guidry21}, to identify the most likely variable objects from over $260{,}000$ high-probability white dwarf candidates in \textit{Gaia} DR2 \citep{GentileFusillo19}. Given the physical similarity of the DAHe objects discovered so far, with masses $0.65{-}0.75\,M_\odot$ and effective temperatures ${\sim}7700$\,K \citep{Gaensicke20}, we then limited the selection to the region of the \textit{Gaia} DR2 Hertzsprung-Russell diagram where DAHe white dwarfs are most likely to reside ($12\!<\!M_G\!<\!14$, $0.2\!<\!G_\text{BP}{-}G_\text{RP}\!<\!0.5$; Figure~\ref{fig:GaiaHRD}). We then collected identification spectra of the highest \textsc{VARINDEX} objects, and, if there were suggestions of DAHe activity, determined a variability period from archival sources or follow-up photometry, and ultimately collected time-series spectroscopy folded on the variability period to produce a complete chronology of the spectral activity. Details of these observations for LP $705{-}64$ and J$1430$ are described in the subsections below. Among our first ${\sim}100$ survey candidates, LP $705{-}64$ and J$1430$ are the first two confirmed DAHe; observations of the remaining objects will be detailed in a future manuscript.
\par LP $705{-}64$ and J$1430$ have \textit{Gaia} DR2 \textsc{VARINDEX} values of $0.0063$ and $0.0131$, respectively; this places LP $705{-}64$ near and J$1430$ well within the top $1\%$ of variable white dwarfs (\textsc{VARINDEX}$_\text{DR2}\!>\!0.0074$; \citealp{Guidry21}). The parallaxes $\varpi$ for these objects are precise enough and suggest a small enough distance ($d\!<\!0.1$ kpc) such that using $d = 1/\varpi$ should provide a sufficiently accurate estimate of the true distance \citep{Luri18}. Using this distance value and the updated \textit{Gaia} DR3 proper motions $\mu_\alpha$ and $\mu_\delta$, we calculate tangential velocity for each object, and find that LP $705{-}64$ has a particularly large $v_\text{tan}$ that is consistent with ${<}30\%$ of low-mass ($0.5{-}0.75\;M_\odot$) and a vanishingly small fraction of intermediate-mass ($0.75{-}0.95\;M_\odot$) white dwarfs \citep{Wegg12}. We discuss the implications of this further in Section~\ref{sec:conc}. These two objects lack sufficient archival survey photometry in the optical and ultraviolet to perform consistent spectral energy distribution fits for $T_\text{eff}$ and $\log g$/mass, so we adopt the atmospheric parameters calculated by \citet{GentileFusillo21b} using \textit{Gaia} photometry and hydrogen-atmosphere white dwarf models with thick ($M_{\rm H}/M_{\rm WD} = 10^{-4}$) hydrogen layers \citep{Kowalski06, Tremblay11}. This collected information is listed in Table~\ref{table:phot}. We note that the use of non-magnetic model atmospheres may make photometrically derived effective temperatures and masses artificially low, as magnetism is known to suppress flux particularly in the \textit{Gaia} $G_\text{BP}$ band, but systematic study suggests this effect is insignificant for stars in the DAHe parameter space \citep{GentileFusillo18,Hardy23}.

\subsection{SOAR/Goodman HTS Identification Spectra}\label{subsec:goodid}
\par Upon sorting our DAHe candidates by \textsc{VARINDEX}, we began collecting identification spectra to detect evidence of spectral activity using the Southern Astrophysical Research (SOAR) 4.1-m telescope and Goodman High-Throughput Spectrograph (HTS) at Cerro Pach\'{o}n, Chile \citep{Clemens04}. We acquired three $300$-second spectra of J$1430$ on 7 April 2021 using a $400$ line mm$^{-1}$ grating and $3.2''$ slit, corresponding to a slit width of $21$ \AA. Our spectral resolution was therefore limited by the wind-impacted observing conditions at a FWHM of $17$ \AA\ ($2.5''$). We bias-subtracted the data and trimmed the overscan regions, then completed reduction using a custom Python routine \citep{Kaiser21}. We flux-calibrated the spectra using standard star EG 274, wavelength-calibrated using HgAr and Ne lamps, and applied a zero-point wavelength correction using sky lines from each exposure. These spectra, when averaged, showed jagged Balmer features suggestive of activity, and we marked J$1430$ for time-series follow-up. 
\par Similarly, we collected five $180$-second spectra of LP $705{-}64$ on 6 August 2021 using the same grating but with a $1''$ ($7$ \AA) slit. Zeeman-split triplet emission features at H$\alpha$ and H$\beta$ were clearly visible in these single spectra, thereby confirming LP $705{-}64$ as a DAHe.

\subsection{\textit{TESS} Photometry}
\par
\begin{figure}
  \includegraphics[width=\columnwidth]{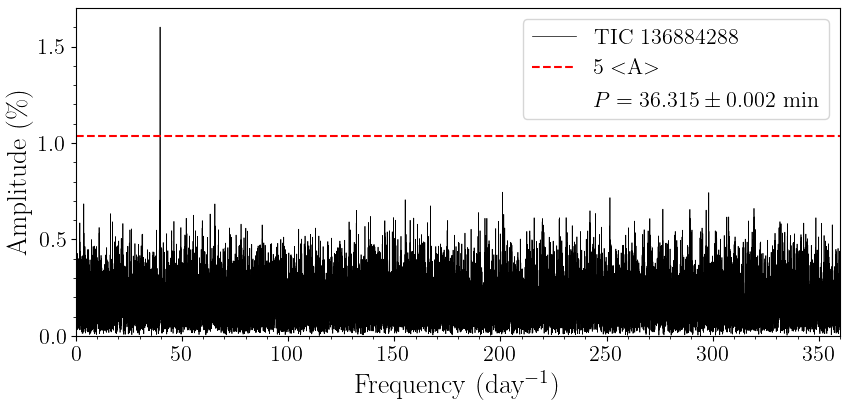}
  \includegraphics[width=\columnwidth]{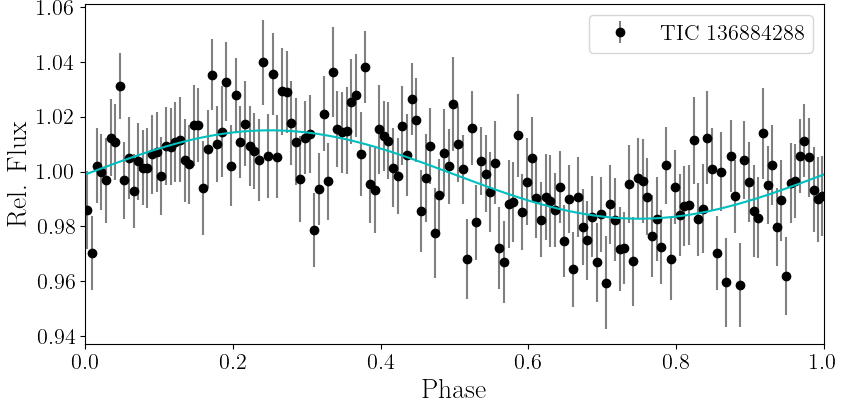}
  \includegraphics[width=\columnwidth]{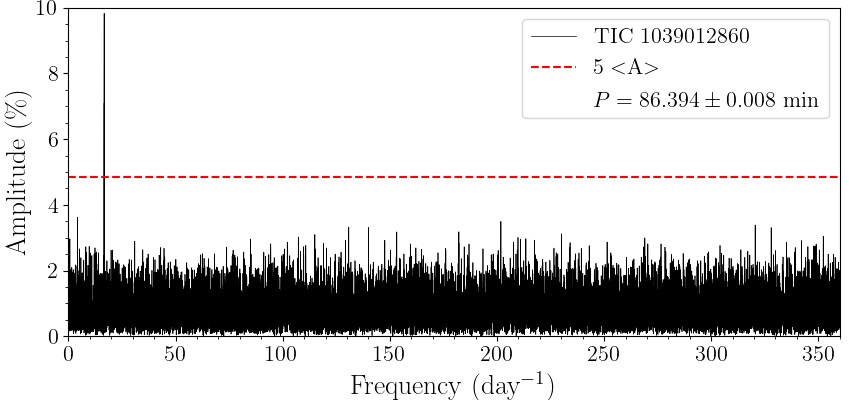}
  \includegraphics[width=\columnwidth]{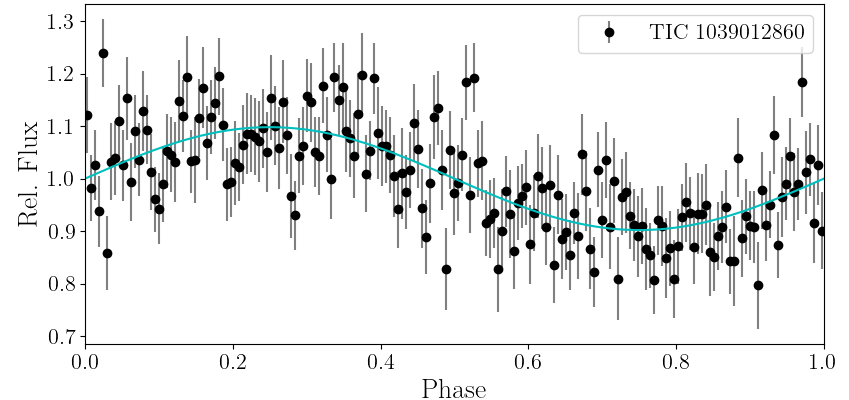}
  \caption{\textit{TESS} periodograms of LP $705{-}64$ (TIC 136884288; top) and J$1430$ (TIC 1039012860; bottom) with a signficance threshold of five times the periodogram average \citep{Baran21}, and phase-folded light curves binned every 100 points. The periodograms suggest dominant photometric periods at $P_\text{LP705-64}\!=\!36.315$ min and $P_\text{J$1430$}\!=\!86.394$ min, respectively, and do not show significant harmonics or additional signals. The likely white dwarf rotation periods are discussed in Section~\ref{subsec:var}.}
  \label{fig:periodogram}
\end{figure}

 The \textit{Transiting Exoplanet Survey Satellite} (\textit{TESS}; \citealp{TESS}) observed LP $705{-}64$ (TIC $136884288$) in Sector 30 with $120$-second exposures, collected from 23 September through 19 October 2020, and J$1430$ (TIC $1039012860$) in Sector 38 with $120$-second exposures, collected from 29 April through 26 May 2021. We extracted these light curves for periodogram analysis using the \textsc{LIGHTKURVE} Python package \citep{LIGHTKURVE}. The periodograms each show one significant peak, whose corresponding periods ($P_\text{LP705-64}\!=\!36.315$ min, $P_\text{J$1430$}\!=\!86.394$ min; Figure~\ref{fig:periodogram}) we adopted for planning our time-series spectroscopy. Later analysis revealed nuance in this variability, which we discuss further in Section~\ref{subsec:var}.

\subsection{SOAR/Goodman HTS Time-Series Spectroscopy}
\par Following our detection of DAHe activity and discernment of photometric variability periods for LP $705{-}64$ and J$1430$, we returned to SOAR and the Goodman HTS to investigate spectral feature variations corresponding to the photometric variability using time-series spectroscopy. On 13 July 2021, we collected $5.4$ hours (approximately four presumed variability cycles) of time-series spectra for J$1430$ in $10$-minute exposures using the same $400$ line mm$^{-1}$ grating and $3.2''$ slit as the identification spectra. The spectra were seeing-limited at $1.5''$, and the average overhead for each acquisition was $5.45$ seconds. We performed the same reductions as were used in the SOAR/Goodman HTS identification spectra (Section~\ref{subsec:goodid}).
\par We targeted LP $705{-}64$ in a similar fashion on 30 August 2021 using $539$-second exposures to reflect equal divisions of the apparent \textit{TESS} variability period, accounting for the overhead time between subsequent exposures. We collected $24$ exposures in this set across $3.6$ hours, corresponding to six presumed variability cycles. We discuss folding and combining spectra in these data sets on divisions of the objects' respective variability periods in Section~\ref{subsec:var}.

\section{Analysis}\label{sec:nlss}

\subsection{Variability and Time-Series Spectroscopy}\label{subsec:var}
\par We performed least squares fits of a sinusoidal signal $A\sin[2\pi(t/P+\phi)]$ to the LP $705{-}64$ and J$1430$ \textit{TESS} light curves using the software \textsc{Period04} \citep{Period04}, where $A$ is the amplitude, $P$ is the variability period, $\phi$ is the phase shift, and $t$ is the observation epoch. Our best-fit value for the period of LP $705{-}64$ is $36.315\pm0.002$ minutes ($f=39.653\pm0.002\,\text{day}^{-1}$), with an amplitude of $1.6\pm0.2\%$, and the best-fit period for J$1430$ is $86.394\pm0.008$ minutes ($f=16.668\pm0.002\,\text{day}^{-1}$), with an amplitude of $9.8\pm0.8\%$.
\par Emulating our process of creating binned spectra for SDSS J$1252{-}0234$ in \citet{Reding20}, we folded our individual spectra of LP $705{-}64$ and J$1430$ into eight equally spaced phase bins, each covering one-eighth of the respective variability periods. We then averaged the exposures within each bin into composite spectra. Our selected exposure time for the LP $705{-}64$ set provided perfect temporal alignment of spectra within each bin, allowing for simple averaging, while for J$1430$ we accounted for blending across phase bins by weighting spectra during rebinning according to the fraction of the acquisition time spanning each bin.
\par The brightnesses of our objects and relatively long exposure times made the Zeeman-split Balmer features visible even in single spectra. For LP $705{-}64$, the folded time-series spectroscopy revealed two distinct emission phases presenting different magnetic field strengths, but which were unexpectedly separated by four ${\sim}9$-minute acquisitions; i.e., one \textit{TESS} period separated the two emission phases, rather than reflecting a full variability cycle. This suggests that a half-rotation, rather than a full rotation, is occurring on this timescale. The true rotation period of LP $705{-}64$ must therefore be twice that of the \textit{TESS} signal at $P_\text{LP705-64}=72.629\pm0.004$ minutes; we have adopted this convention throughout. Our other target, J$1430$, returned to its original orientation on the same period as the \textit{TESS} signal, so we infer its rotation period to be $P_\text{LP705-64}=86.394\pm0.008$ minutes.
\par Past DAHe discoveries all present maximal emission at photometric minimum, and LP $705{-}64$ and J$1430$ appear to follow this same trend by visual inspection of the slopes of spectral continua---the emission phases are present when the continua have the flattest slopes. However, for both of our objects, these slopes eventually become unreliable due to encroaching clouds in the final few exposures. Consequently, we do not attempt to convolve our binned spectra of LP $705{-}64$ and J$1430$ with theoretical filter profiles to obtain rough ``light curves'' of our acquisitions, as we did for SDSS J$1252{-}0234$ in \citet{Reding20}. We also did not select acquisition times based on anticipated photometric variability phases projected from \textit{TESS} ephemerides, as our time-series spectroscopy was too far separated in time from the \textit{TESS} photometry to predict times of maxima or minima with sufficient accuracy. Instead, our division of the variability periods into eight bins provides enough temporal resolution to select spectral phases close to expected photometric maxima and minima.
 
{\renewcommand{\arraystretch}{1}
\begin{table*}
  \centering
  \caption{LP $705{-}64$ and J$1430$ magnetic field strengths as measured from H$\alpha$ and H$\beta$ at notable spectral phases (Figure~\ref{fig:mag}). Field strength estimates are derived from least squares Gaussian profile component fits and magnetic transitions calculated in \citet{Schimeczek14}.}\label{table:fit}
  \begin{threeparttable}
    \begin{tabular*}{1.31\columnwidth}{l l l | r r | r}
      \hline
      Object & Phase & $T (\text{BMJD}_\text{TDB})$ & $B\text{ (MG), H}\beta$ & $B\text{ (MG), H}\alpha$ & B (MG), Avg.\\
      \hline
      LP $705{-}64$ & Em. Wide & $59457.182074$ & - & $22.2\pm0.9$ & $22.2\pm0.9$\\
      & Em. Narrow & $59457.207290$ & $9.3\pm1.0$ & $9.5\pm0.8$ & $9.4\pm0.6$\\
      \hline
      J$1430$ & Emission & $59409.011568$ & $9.3\pm0.1$\tnote{*} & $8.6\pm1.4$ & $8.9\pm1.4$\\
      & Absorption & $59409.041564$ & $5.8\pm0.3$ & - & $5.8\pm0.3$\\
      \hline
    \end{tabular*}
    \begin{tablenotes}
      \item[*] This is a single measurement from $\sigma_-$ as the only component fully visible in this phase. We weight the average $B$ for this phase accordingly.
    \end{tablenotes}
    \end{threeparttable}
\end{table*}
}

\begin{figure*}
\includegraphics[width=0.475\textwidth]{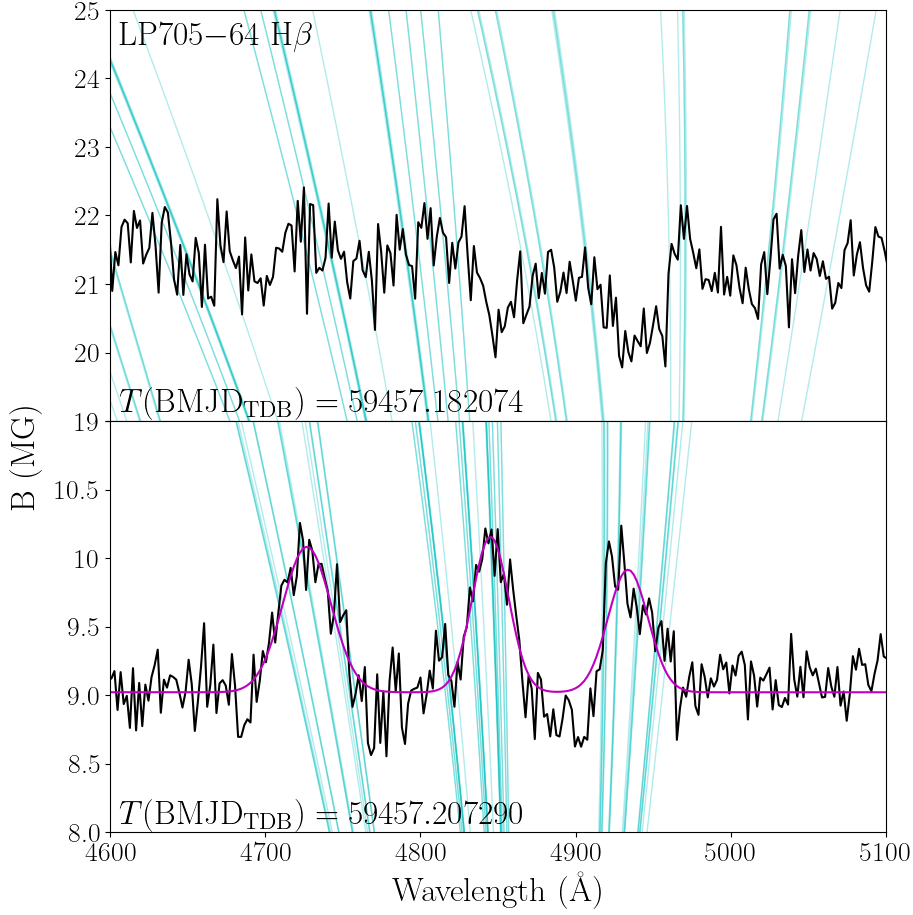}\includegraphics[width=0.475\textwidth]{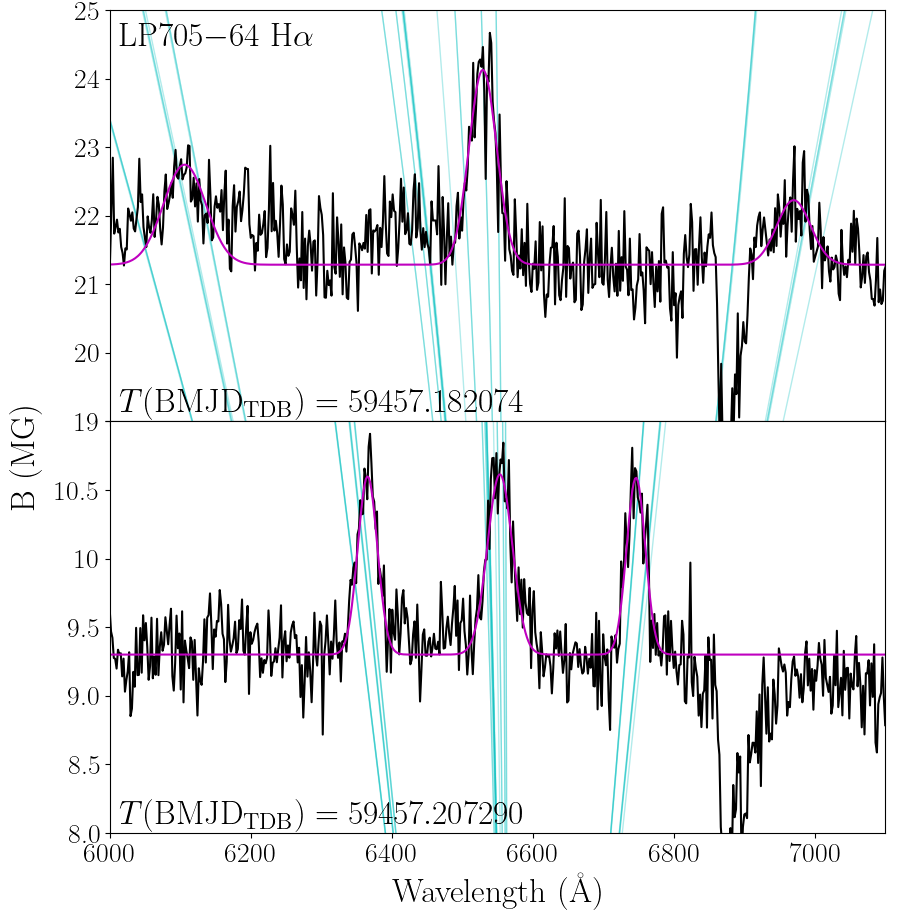}
\includegraphics[width=0.475\textwidth]{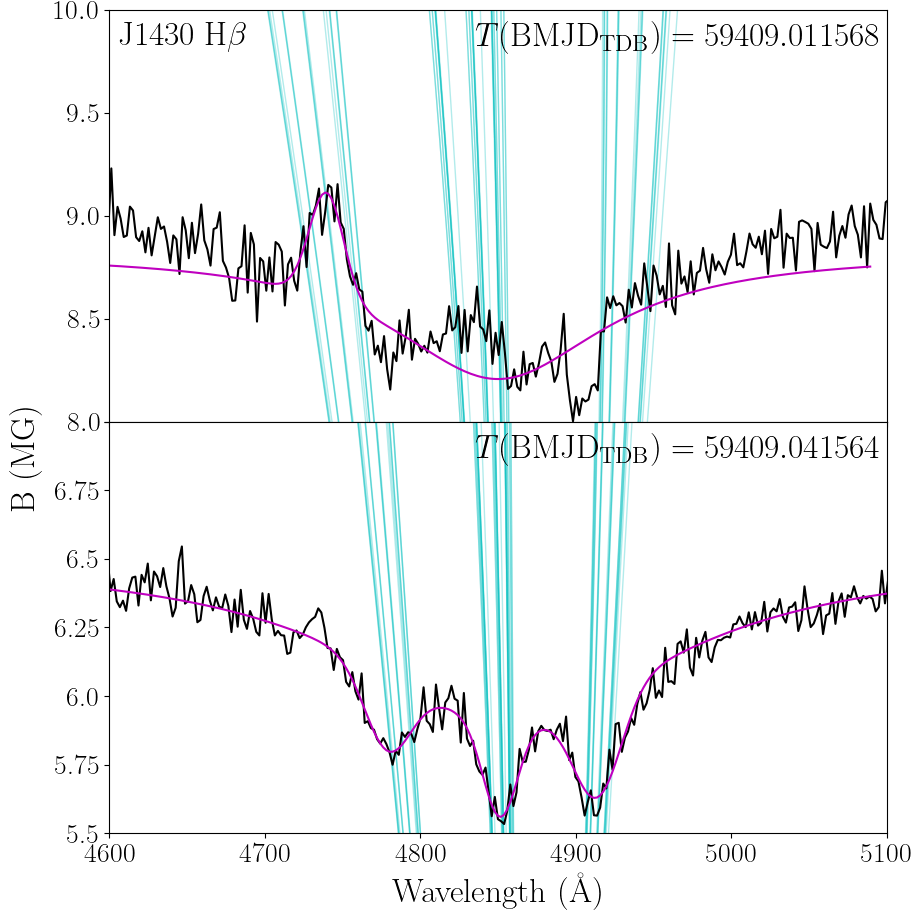}\includegraphics[width=0.475\textwidth]{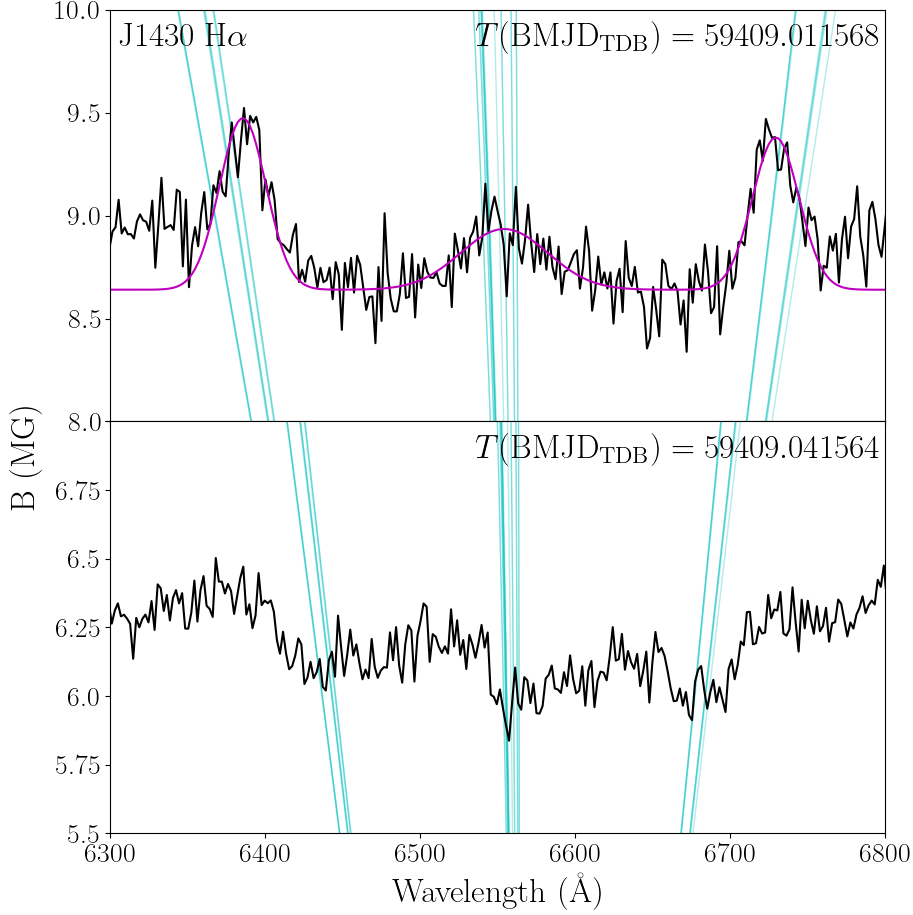}
  \caption{H$\beta$ and H$\alpha$ profiles of LP $705{-}64$ and J$1430$ from binned SOAR spectra at maximum emission and absorption phases. We determine feature locations from least squares fitting where possible, and report corresponding magnetic field strengths in Table~\ref{table:fit}. The absorption feature at ${\sim}6900$\,\AA\ in the H$\alpha$ spectra of LP $705{-}64$ (b) is telluric, which partially obscures the $\sigma_+$ Zeeman component in the wide emission phase. Given the uncertainty in projecting the \textit{TESS} ephemerides forward to our spectral acquisition times, we instead present the midpoints of the binned spectra in BMJD$_\text{TDB}$ as markers for future analysis. The data behind this figure are available in the article's online supplementary material.}
  \label{fig:mag}
\end{figure*}

\par For LP $705{-}64$, the maximum magnetic field strength visible in the top panels of Figure~\ref{fig:mag} occurs at a phase within 5\% of the photometric minimum of the \textit{TESS} observations. The maximum field strength (emission phase) for J$1430$ also occurs significantly closer to the \textit{TESS} photometric minimum than the weaker magnetic absorption phase. However, extrapolating the ephemeris uncertainties forward to our spectral acquisition times produces error bars on these associations that span nearly a full variability cycle. We therefore invite additional photometric observations that can better reveal the light curve morphology and confidently associate the notable spectral phases with maxima and minima.

\subsection{Magnetic Field Strengths}
\par To determine magnetic field strengths, we performed least squares fits of the H$\alpha$ and H$\beta$ profiles at maximum emission and absorption phases using the Python package \textsc{LMFIT} \citep{Newville14}. We used a Lorentzian profile for wide absorption features, where applicable, and Gaussian profiles for individual Zeeman components. After finding centroid locations of the feature components, we converted these into magnetic field strength estimates using the magnetic transitions catalogued in \citet{Schimeczek14}. Unlike previously analyzed DAHe, both LP $705{-}64$ and J$1430$ host magnetic fields that evolve significantly across their rotational periods (Figure~\ref{fig:mag}).
\par For LP $705{-}64$, H$\alpha$ and H$\beta$ manifest as Zeeman-split emission with no underlying absorption in both notable spectral phases. In the narrower triplet emission phase, both features are consistent with a field strength of $B\!=\!9.4\pm0.6$ MG, before they disappear into the continuum and reappear as significantly wider Zeeman emission corresponding to a field strength of $B\!=\!22.2\pm0.9$ MG. These values represent the weighted averages of the individual field strength estimates from each Zeeman component, with weights determined by respective uncertainties. The overall uncertainty on the weighted average is the standard deviation of the maximally dispersed component estimates. 
\par J$1430$ mimics SDSS J$1252{-}0234$ in exhibiting H$\beta$ absorption at apparent photometric maximum, and emission at photometric minimum. However, unlike its predecessor, in J$1430$ this absorption is Zeeman-split with a magnetic field strength of $B\!=\!5.8\pm0.3$ MG, which becomes partially and asymmetrically filled by emission from a stronger magnetic field of $B\!=\!8.9\pm1.4$ MG at the opposite phase. H$\alpha$ more prominently displays this emission as a fully resolved triplet, allowing for easy calculation of this field strength.

\subsection{Magnetic Field Geometries}\label{subsec:vard}
\par \citet{Reding20} found that SDSS J$1252{-}0234$ transitions across its variability period from presenting slightly broadened, but not Zeeman-split, Balmer absorption features at photometric maximum, to revealing its significant H$\beta$ Zeeman triplet emission at photometric minimum. This indicates that the magnetic spot, above which the emission manifests and displays the strongest concentration of magnetic field lines, is oriented along our line of sight at photometric minimum. This orientation consequently provides the best measure of the polar magnetic field, while the absorption center in the opposite phase returns closer to the rest-frame wavelength of H$\beta$. This evolution of the apparent magnetic influence on Balmer features with stellar rotation illustrates that we observe different hemisphere-averaged magnetic fields across the variability phases.
\par J$1430$ behaves similarly in transitioning from absorption to emission with an apparent growing magnetic field strength, but differs in its strongly Zeeman-split absorption phase. Furthermore, the J$1430$ emission phase does not seem to show a clean single-field feature as was seen in SDSS J$1252{-}0234$; rather, the previous Zeeman absorption seems to still be present and asymetrically filled, possibly indicating a superposition of two apparent magnetic field signatures. H$\alpha$, conversely, does not show as complicated a transition, with its emission phase presenting an easily measurable feature corresponding to a single field.
\par LP $705{-}64$ adds further complexity in magnetic field presentation by never showing absorption, but instead exhibiting two emission phases corresponding to drastically different magnetic field strengths at its photometric minima. These two new discoveries therefore break the previous DAHe mold by presenting multiple distinct magnetic field signatures, while the previous three members only displayed Zeeman splitting corresponding to single field strengths. However, as with other known magnetic white dwarfs (e.g., \citealt{Martin84}), it is often difficult to distinguish offset dipole emission from higher-order field geometries.

\subsection{Companion Limits}
\par Owing to their small sizes, the only potential binary systems that can fit within non-overluminous apparently isolated white dwarf spectral energy distributions are double degenerate systems containing at least one white dwarf of very high mass, or substellar companions which emit most strongly in infrared wavebands. The former case invokes a super-Chandrasekhar-mass binary system, which has never been observed even in targeted searches for the most extreme supernova Ia progenitors \citep{RebassaMansergas19}. We disregard this for LP $705{-}64$ and J$1430$ given the lack of substantial radial velocity variability, and only assess potential substellar companions.
\par The \textit{Wide-field Infrared Survey Explorer} (\textit{WISE}; \citealp{Wright10}) collected infrared photometry of both LP $705{-}64$ (WISE J$003513.00{-}122510.8$) and J$1430$ (WISE J$143019.77{-}562357.5$) in 2015, which was reported in the CatWISE2020 catalog \citep{Marocco21}. We use these measurements and the averaged \textit{WISE} photometry for late-spectral-type objects from the Database of Ultracool Parallaxes \citep{Dupuy12} to place limits on potential substellar companions to each white dwarf. Because the peak wavelengths of these substellar objects fall in the far-infrared, their fluxes typically rise when moving from the $W1$ to $W2$ bands, which runs opposite to the declining trend seen in white dwarfs whose peak wavelengths push into the ultraviolet. The $W2$ band therefore places the strongest constraints on companion spectral type. We find that $W2$ photometry of spectral type T4 exceeds the corresponding point for LP $705{-}64$ by over $3\text{-}\sigma$, while spectral type T2 similarly exceeds the $W2$ photometry of J$1430$. We therefore rule out a stellar or substellar companion earlier than spectral type T.

\section{Discussion and Conclusions}\label{sec:conc}
\par We present the discoveries of two new DAHe white dwarfs, LP $705{-}64$ ($0.81\pm0.04\,M_\odot$, $T_\text{eff}=8440\pm200$ K) and WD J$143019.29{-}562358.33$ ($0.83\pm0.03\,M_\odot$, $T_\text{eff}=8500\pm170$ K), bringing the total population of the DAHe class to five objects\footnote{After receipt of this paper's referee report, a preprint was posted announcing spectroscopic identification of 21 northern-hemisphere DAHe white dwarfs from the Dark Energy Spectroscopic Instrument (DESI) survey \citep{Manser23}.}.
\par Using time-series spectroscopy from the $4.1$-m SOAR Telescope, we captured signatures of evolving magnetic fields in each star with rotational phase, setting them apart from the previously discovered members of this class which only presented Zeeman splitting corresponding to single magnetic field strengths. LP $705{-}64$ appears to rotate at $72.629$ minutes and displays Zeeman-split Balmer emission at two separate emission phases, corresponding to magnetic field strengths of $B\!=\!9.4\pm0.6$ MG and $B\!=\!22.2\pm0.9$ MG. At its weakest, J$1430$ presents Zeeman-split Balmer absorption corresponding to a magnetic field strength of $B\!=\!5.8\pm0.3$ MG. Half an $86.394$-minute rotation cycle later, the absorption appears superimposed with Balmer emission corresponding to a stronger magnetic field strength of $B\!=\!8.9\pm1.4$ MG. As with the previously known DAHe, the maximum magnetic field strength appears coincident with the photometric minimum from the TESS observations, although phasing over a many-months baseline carries uncertainty. In the case of LP $705{-}64$ we have shown that DAHe white dwarfs can show two magnetic poles.
\par With five members now known, the DAHe class remains relatively homogenous in its physical characteristics, with all members having mega-gauss magnetic fields, effective temperatures $7500{-}8500$ K, and masses slightly higher than but near the white dwarf population average of $0.62\,M_\odot$ \citep{GenestBeaulieu19}. The nature of the mechanism driving their emission remains elusive, however, as all are apparently isolated with no detectable stellar companion. As pointed out by \citet{Walters21}, the physical similarities of known DAHe strongly suggest the variability mechanism is not extrinsic, and likely represents a phase of evolution for at least some white dwarf stars.
\par Thus, the origin of DAHe white dwarfs remains a mystery. In addition to strong magnetism, most of the known DAHe rotate significantly faster than a typical white dwarf. One way to generate strong magnetism and rapid rotation in white dwarfs is via a past stellar merger, especially of two white dwarfs \citep{Ferrario97, Tout08, Nordhaus11}. Double-degenerate mergers may produce more massive white dwarfs, but the merger of two low-mass white dwarfs (${\lesssim}0.4\,M_\odot$) can produce a single remnant with a mass near the white dwarf average \citep{Dan14}. It has also been speculated that planetary engulfment may spin up white dwarfs enough to generate magnetic dynamo activity \citep{Kawka19, Schreiber21b}. 
\par Another indicator of a merger origin is a mismatch between expected cooling age and apparent age as inferred from kinematics. This reasoning was used to classify hot carbon-atmosphere (DQ) white dwarfs as likely merger products \citep{Dunlap15}. If descended from single stars without external interactions, initial-final mass relations and cooling models suggest that the DAHe white dwarfs should have ${\sim}3\,M_\odot$ progenitors and roughly $2$-Gyr total ages \citep{Cummings18, Bedard20}. Kinematic outliers could help reveal if any DAHe are merger byproducts, though this is best performed on a population rather than a single object \citep{Cheng20}. In this context, the relatively fast kinematics of LP $705{-}64$ ($v_\text{tan}=52.89$ km s$^{-1}$) are interesting, although are not necessarily direct evidence of a past interaction. The other known DAHe have relatively slow kinematics, with $v_\text{tan}\approx5{-}35$ km s$^{-1}$. Discovery and analysis of a larger sample of DAHe will better inform the kinematic ages of this sample.

\section*{acknowledgements}This work is based on observations obtained at the Southern Astrophysical Research (SOAR) telescope, which is a joint project of the Minist\'{e}rio da Ci\^{e}ncia, Tecnologia, Inova\c{c}\~{o}es e Comunica\c{c}\~{o}es (MCTIC) do Brasil, the U.S. National Optical Astronomy Observatory (NOAO), the University of North Carolina at Chapel Hill (UNC), and Michigan State University (MSU). Support for this work was in part provided by NASA TESS Cycle 2 Grant 80NSSC20K0592 and Cycle 4 grant 80NSSC22K0737, as well as the National Science Foundation under grant No. NSF PHY-1748958. We acknowledge NOIRLab programs SOAR2021B-007 and SOAR2022A-005, as well as excellent support from the SOAR AEON telescope operators, especially C\'esar Brice\~no. Some of the data presented in this paper were obtained from the Mikulski Archive for Space Telescopes (MAST). STScI is operated by the Association of Universities for Research in Astronomy, Inc., under NASA contract NAS5-26555. Support for MAST for non-HST data is provided by the NASA Office of Space Science via grant NNX13AC07G and by other grants and contracts. The \textit{TESS} data may be obtained from the MAST archive (Observation ID: tess2020266004630-s0030-0000000136884288-0195-s). This research made use of Lightkurve, a Python package for Kepler and TESS data analysis \citep{LIGHTKURVE}. This work has made use of data from the European Space Agency (ESA) mission {\it Gaia} (\url{https://www.cosmos.esa.int/gaia}), processed by the {\it Gaia} Data Processing and Analysis Consortium (DPAC, \url{https://www.cosmos.esa.int/web/gaia/dpac/consortium}). Funding for the DPAC has been provided by national institutions, in particular the institutions participating in the {\it Gaia} Multilateral Agreement. This work makes use of data products from the Wide-field Infrared Survey Explorer, which is a joint project of the University of California, Los Angeles, and the Jet Propulsion Laboratory/California Institute of Technology, funded by the National Aeronautics and Space Administration. This research has made use of NASA's Astrophysics Data System. This research has made use of the VizieR catalogue access tool, CDS, Strasbourg, France. This research has made use of the SIMBAD database, operated at CDS, Strasbourg, France. This research made use of Astropy, a community-developed core Python package for Astronomy \citep{ASTROPY1, ASTROPY2}. This research made use of SciPy \citep{SCIPY}. This research made use of NumPy \citep{NUMPY}. This research made use of matplotlib, a Python library for publication quality graphics \citep{MATPLOTLIB}. This work made use of the IPython package \citep{IPYTHON}. 


\bibliographystyle{mnras}
\bibliography{/home/jsreding/Documents/Work_Education/UNC/Research/references.bib}

\section*{Data Availability Statement} The data underlying this article are available in the article and in its online supplementary material. 


\bsp	
\label{lastpage}

\end{document}